# Impact of solar wind disappearance event on the Martian nightside ionospheric species: First results

L. Ram[1], D. Rout[3], S. Sarkhel[1,2,*]

*Sumanta Sarkhel, Department of Physics, Indian Institute of Technology Roorkee, Roorkee - 247667, Uttarakhand, India (sarkhel@ph.iitr.ac.in)

[1]Department of Physics,
Indian Institute of Technology Roorkee,
Roorkee - 247667
Uttarakhand, India

[2]Centre for Space Science and Technology,
Indian Institute of Technology Roorkee,
Roorkee - 247667
Uttarakhand, India

[3]National Atmospheric Research Laboratory,
Gadanki, India

**Abstract**

The impact of a rarest solar wind phenomenon [disappearing solar wind (DSW) event during 26-28 December 2022] on the Martian nightside ionosphere is investigated, for the first time, using MAVEN datasets. During an extremely low solar wind density, the Martian top nightside ionospheric species underwent significant enhancements. At a given altitude, the density of electron increased by ~2.5 times, while for ions ($NO^+$, $O_2^+$, $CO_2^+$, $C^+$, $N^+$, $O^+$, and $OH^+$), it increased by >10 times, respectively, compared to their typical average quiet-time periods. The extremely low dynamic pressure during the DSW event, contrasting with relatively higher nightside ionospheric thermal pressure nearly by an order, suggests an upward ionospheric expansion as a direct consequence. This causes an increased topside ionospheric density. Additionally, the day-to-night plasma transport may also be a contributing factor to increased plasma density. Thus, this study presents a new insight about planetary atmosphere/ionosphere response to the rarest event.

**Key Points:**
- A significant increase in the Martian nightside ionospheric electron and ions density is observed during the solar wind disappearance event.

- At a given altitude, the electron and ions densities increased by a factor of ~2.5 and >10 respectively, compared to the average quiet-time.

- An extremely low dynamic pressure compared to higher ionospheric thermal pressure on the nightside causes increased plasma densities.

**Plain Language Summary**

The solar wind-Mars interactions play a pivotal role in the evolution of Mars climate over time. Apart from Earth, there is a vague understanding of the influence of extremely low density and pressure solar wind event i.e., disappearing solar wind (DSW), on the planetary atmosphere/ionosphere, specifically on our neighboring planet, Mars. In order to understand this aspect, we have explored the behavior of Martian nightside ionospheric species i.e., $e^-$, $NO^+$, $O_2^+$, $CO_2^+$, $C^+$, $N^+$, $O^+$, and $OH^+$ during the DSW event. An increase in the densities of ionospheric species is observed on the topside ionosphere. At a given altitude, the nightside ionospheric densities are ~2.5 (for electrons) and more than 10 (for ions) times higher in magnitude compared to their typical average quiet-time periods. This indicates an expansion from the lower to the topside ionosphere, majorly caused by the higher ionospheric thermal pressure as compared to the external low solar wind dynamic pressure during the DSW. Furthermore, the plasma transport from dayside to nightside may also increase the ionospheric density in the nightside. Hence, this study, for the first instance, guides us to a new understanding of the impact of a rarest solar wind phenomenon on the planetary atmosphere/ionosphere.

# 1. Introduction

The solar wind, originating from the solar corona, extends radially outward throughout our solar system, carrying the interplanetary magnetic field (IMF) within its flow. As it travels in space, the solar wind interacts with various celestial bodies, including planets, moons, asteroids, and comets. The nature of this interaction is primarily influenced by the intrinsic magnetic field and the atmospheric conditions of these bodies. For instance, the solar wind interacts with the Earth's magnetosphere due to the presence of the planet's intrinsic dipolar magnetic field. In contrast, on Venus, lacking a significant magnetic field, the solar wind interacts directly with the ionosphere. In the present context, Mars is an interesting candidate in the solar system as it possesses a localized crustal magnetic field which results in both magnetized and unmagnetized magnetospheres and acts as a hybrid magnetosphere (Acuna et al., 1999; Fowler et al., 2024).

During the period 26-28 December 2022, the Mars Atmosphere and Volatile EvolutioN (MAVEN) spacecraft (Jakosky et al., 2015) observed an extremely low solar wind density stream (<1 cm$^{-3}$) for an extended period (>24 h) at Mars which was first encountered by Earth. This unusual and extended density depletion was also accompanied by very low-velocity solar wind flows (<250 km s$^{-1}$), which resulted in depletion in the solar wind dynamic pressure. This caused a dramatic expansion of the Martian magnetosphere and bow shock (Halekas et al., 2023). Estimations suggest that the boundaries of the Martian magnetosphere expanded by thousands of kilometers, surpassing the typical location of the bow shock (Halekas et al., 2023). This remarkable event is famously known as "The Day the Solar Wind Disappeared at Mars" or disappearing solar wind (DSW) (Halekas et al., 2023; Shaver et al., 2024; Xu et al., 2024). A similar type of low-density event was first detected by the Advanced Composition Explorer (ACE) satellite on Earth during 10-12 May 1999 (Lockwood, 2001; Janardhan et al., 2008). Recently, it has been revealed that the origin of such solar wind disappearance events is mostly associated with CIR events (Rout et al., 2023; Shaver et al. 2024). The study of Rout et al. (2023) showed the relation between these low-density events and solar wind non-radial flow and their impact on Earth's magnetosphere. The impact of this event on the Earth's magnetosphere-ionosphere system has also been reported by various studies (Farrugia et al., 2008; Le et al., 2000; Ohtani et al., 2000). However, it is particularly noteworthy to mention the observation of a similar rare event at Mars during a fortuitous MAVEN spacecraft orbit. This unique opportunity enables us to further explore its impact on the Martian ionosphere.

Few studies have delved into the interaction between the solar wind and Mars under tenuous solar wind conditions, as well as its impact on Mars's magnetosphere-ionosphere system (Fowler et al., 2024; Halekas et al., 2023; Shaver et al., 2024; Xu et al., 2024). Notably,

Xu et al. (2024) explored the dynamics of superthermal electrons on Mars during a solar wind disappearance event. During this event, the ionosphere experienced expansion, and the upstream solar wind exhibited the ability to alter the Mars-solar wind interaction state within remarkably short timeframes-less than one MAVEN orbit (~3.5 hours) (Fowler et al., 2024). However, no study has yet investigated the impact of this extremely low-density event on ionospheric species and plasma density. Thus, this study, for the first time, reports the effects on ionospheric species for both electrons and ions species, i.e., heavier ($NO^+$, $O_2^+$, and $CO_2^+$) and lighter ($C^+$, $N^+$, $O^+$, and $OH^+$) ions on the nightside. This event presents a unique opportunity to probe planetary interactions' dynamic and evolving nature amidst tenuous solar wind conditions.

## 2. Data

The low-density solar wind event was observed by a MAVEN spacecraft on 26-28 December 2022 on Mars. In order to address the variation of Martian ionospheric species during this event, we utilized the datasets from multi-instruments aboard MAVEN for observations of the solar wind, IMF, and Martian plasma variation. During December 2022, the MAVEN spacecraft followed an elliptical orbit of nearly 3.5 hours with an apoapsis and a periapsis around 4500 and 180-190 km altitudes, respectively. The MAVEN periapsis surveils the nightside ionosphere over southern latitudes (~0-70° S) during this period. In the present study, the sampling of the upstream solar wind and IMF conditions are performed by leveraging the in-situ key parameter datasets of Solar Sind Ion Analyzer (SWIA; Dunn, 2023; Halekas et al., 2015) and Magnetometer (MAG; Connerney, 2015; Dunn, 2023). The SWIA instrument is a cylindrical symmetric electrostatic analyzer that measures ion velocity distribution in the 3-axis and covers a broad energy range of 5-25000 eV. It monitors the plasma in a 360 × 90° field of view with a resolution of ~22.5° (Halekas et al., 2015). Measurements of MAG instrument have been used to calculate the resultant interplanetary magnetic field (IMF (|**B**|) over the period of observation. The MAVEN MAG instrument is a tri-axial fluxgate magnetometer that measures the ambient vector magnetic field in three dimensions during spacecraft orbit and covers a range of ±65536 nT. The SWIA and MAG in-situ key parameters datasets were accessed and analyzed using the Python Data Analysis and Visualization tool (PyDIVIDE; MAVEN SDC et al., 2020) and NASA Planetary Data System (PDS).

The measurements of electron density were obtained using the Langmuir Probe and Waves (LPW; Andersson et al., 2015; Dunn, 2023) instrument onboard MAVEN. The LPW instrument is optimized to probe the electron density falls within the range of $10^2$-$10^6$ cm$^{-3}$. The data is accessed using the PyDIVIDE tool. In addition, the Martian nightside ionospheric species densities data was obtained using a Neutral Gas and Ion Mass Spectrometer (NGIMS; Mahaffy, Benna et al., 2015) aboard MAVEN. The NGIMS instrument is designed to measure the ions and neutrals in the mass range of 2-150 amu with unit resolution and presently monitors the Martian ionosphere between 180 and 500 km altitudes during MAVEN spacecraft periapsis. The NGIMS datasets were accessed using the MAVEN science data center (SDC) and NASA PDS. The information related to different levels and versions of the datasets associated with SWIA, MAG, LPW, and NGIMS instruments is provided in the data availability section below.

## 3. Results

On 26 December 2022, an extremely low-density solar wind was observed on Mars (Halekas et al., 2023). In order to investigate the Martian ionospheric response to this low-density solar wind event, we have explored the topside ionosphere in the nightside. The following sub-sections show the results on the variation of solar wind and the Martian ionospheric species during the DSW event in comparison to the pre- and post-event periods.

### 3.1 Observations of solar wind and IMF on Mars from 15-31 December 2022 using SWIA and MAG datasets onboard MAVEN

Fig. 1 depicts the upstream solar wind plasma and IMF observations between 15 and 31 December 2022. For the pristine solar wind measurements, we leverage the Halekas et al. (2017) algorithm above 4400 km altitude from Mars surface. The *y*-axis for each panel (Fig. 1) from top to bottom (a-d) represents the solar wind density, velocity, solar wind dynamic pressure (SWDP), and resultant IMF ($|\mathbf{B}|$) variation with respect to the respective days of December 2022, marked on the *x*-axis. The observations of solar wind and IMF indicate a corotating interaction region (CIR) event that started on 21 December and finally ended on 25 December with maximum enhancement in velocity (Shaver et al., 2024). The CIR passage is followed by a rarefaction region with very low solar wind density from 26-28 December 2022. In order to find the Martian ionospheric response during this extremely low solar wind density event, we have analyzed the Martian ionospheric species for pre-, post-, and during the DSW event. We selected the pre- and post-event datasets outside the CIR to provide a quiet background scenario, enabling a better comparison with DSW event. The vertical dotted lines illustrate MAVEN orbits during the pre-event (black color), DSW event (red color), and post-event (magenta color). The vertically dotted colored lines indicate the Martian ionospheric species profile variation that is illustrated with similar color scheme in the next section. During the DSW event, the solar wind density and dynamic pressure almost disappeared and approached nearly 0.15 cm$^{-3}$ and 0.014 nPa, respectively (Fig. 1a&c). Fig. 1b shows the decline in the solar wind velocity during DSW with the lowest magnitude approaching nearly 228 km

s$^{-1}$. The resultant IMF (|**B**|) during this period lies near to 7 nT (Fig. 1d). Due to MAVEN spacecraft trajectory constraints and elliptical orbit, it does not take the upstream solar wind and ionosphere measurements simultaneously and having a difference of a few hours (Jakosky et al., 2015). All selected profiles (shown by vertically dotted colored lines) during DSW, pre- & post-event periods fell within the similar SZA (~120-170°) range, in the deep nightside.

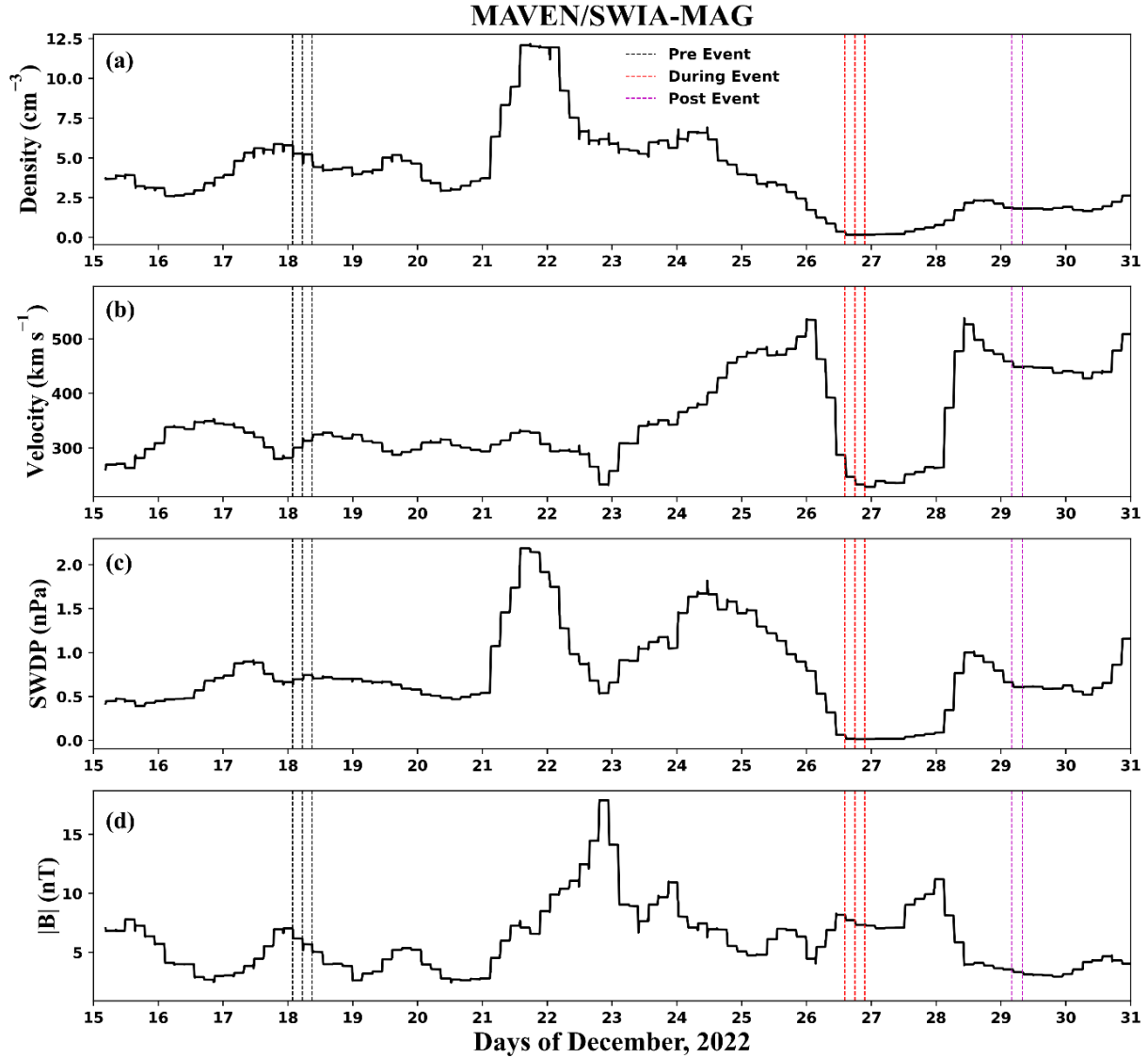

**Fig. 1.** The MAVEN/SWIA-MAG Observations of the solar wind (a-d) density, velocity, dynamic pressure, and IMF (|**B**|) respectively near Mars, during December 2022.

## 3.2 Variation of the Martian ionospheric species in the nightside during the DSW event on 26 December 2022

In order to address the DSW event's impact on the Martian topside ionosphere, we have utilized the LPW and NGIMS measurements onboard the MAVEN spacecraft. The MAVEN spacecraft measures the Martian ionospheric composition during periapsis (Andersson et al., 2015; Benna et al., 2015; Jakosky et al., 2015). During December 2022, the MAVEN inbound phase, monitors the ionospheric compositions primarily on the nightside of Mars' southern latitudes. However, during the MAVEN outbound leg, the dayside measurements are falling at higher altitudes with different SZA, primarily deprived of ionospheric data and cannot be utilized for the ionospheric study. Therefore, in the present work, we have primarily presented and focused on the behavior of the Martian nightside ionosphere in response to the DSW event.

Fig. 2. represents the LPW and NGIMS in-situ measurements for average electron and ion density as a function of altitude. A similar color scheme has been followed as shown in Fig. 1 for pre-, during, and post-event profiles. These profiles were calculated by averaging a 7 km altitude bin of density (shown by black, red, and magenta colors in Fig. 2 and at least 2-3 profiles with similar SZA range have considered). Due to the unavailability of complete density profiles for electron and ions on 27 December 2022, only the profiles from 26 December 2022 (red dotted vertical lines in Fig. 1), along with their corresponding standard deviation bars (red color), are shown during the DSW event. The top panel of Fig. 2 represents the variation of the electron and heavier ions ($NO^+$, $O_2^+$, and $CO_2^+$), whereas, the bottom panel represents the variation of lighter ions ($C^+$, $N^+$, $O^+$, and $OH^+$) with respect to the altitude. Fig. 2 shows a noticeable change in the magnitude of electrons, heavier, and lighter ions density during the DSW event compared to the pre- and post-event. During the DSW event (red color), the electron density (Fig. 2a) increases between 200 and 280 km altitudes with maximum magnitude occurring at ~250 km altitude compared to pre- and post-event (black and magenta colors) periods. During the DSW event, the increased electron density around 250 km is ~2.5 times higher than average quiet time magnitude. Similarly, for heavier and lighter ions (Fig. 2b-h), the increased density profiles are observed during DSW event as compared to the pre- and post-event periods. The higher magnitude of ion density is observed at 235 and 280 km altitudes, respectively during the DSW, whereas for pre-and post-event scenario, it occurs at altitude range of 195-230 km. During the event, the maximum densities for heavier ions i.e., $NO^+$, $O_2^+$ and $CO_2^+$ are respectively 18, 39, and 21 times higher, whereas, for lighter ions i.e., $C^+$, $N^+$, $O^+$, and $OH^+$, the magnitude are respectively 15, 10, 67, and 23 times higher as compared to their average quiet-time periods at a given altitude. In addition, Fig. 2 also depicts the highly variable ionospheric profiles above 200 km as depicted by 1σ deviation during the DSW event in comparison to the pre- and post-event scenarios. Although, the large standard deviation of pre-and post-event profiles near periapsis (~180-190 km) is primarily due to MAVEN maneuvering from inbound to outbound. In totality, during the DSW event, the behavior of the ion density profiles is quite similar for both heavier and lighter ions in the topside ionosphere. For brevity, the detailed information related to the maximum magnitude of average electron and ions densities with respect to altitude during pre-, DSW, and post-event is provided in Table 1.

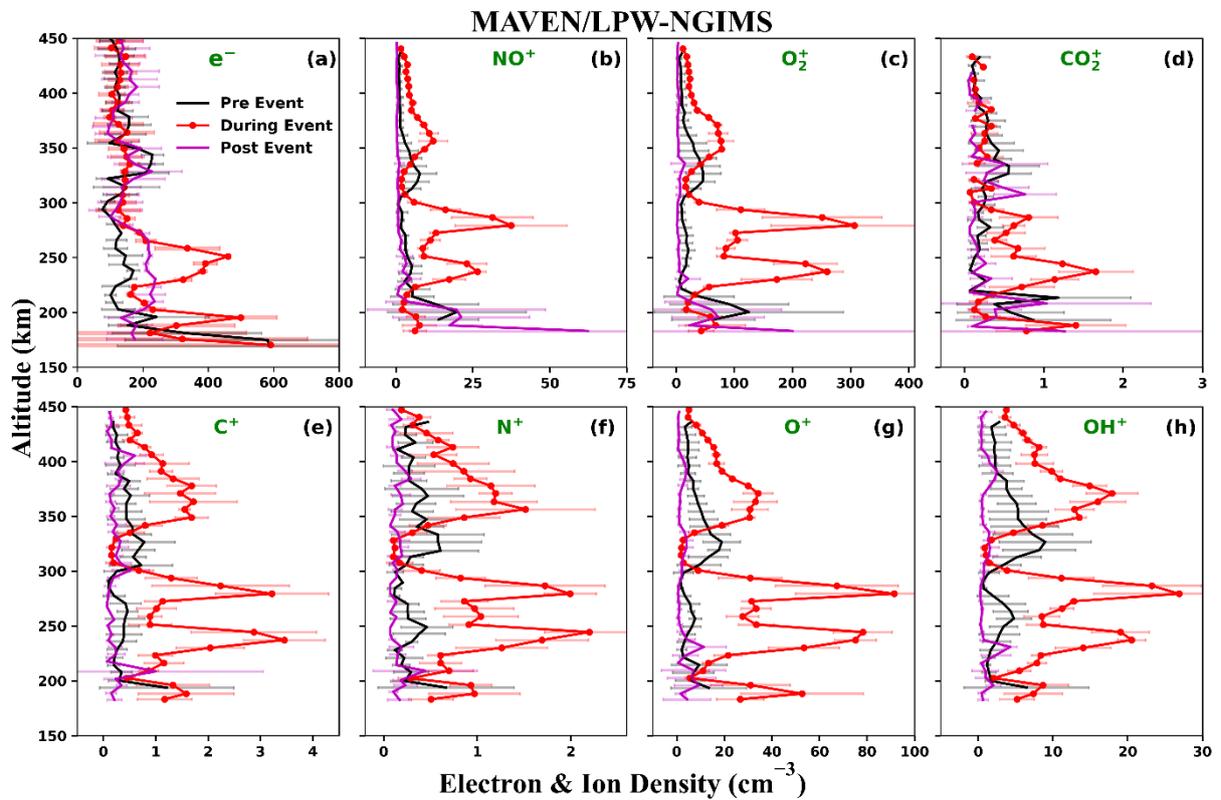

**Fig. 2.** The Langmuir Probe & Wave (LPW) and Neutral Gas and Ion Mass Spectrometer (NGIMS), observations for Martian nightside ionospheric species, i.e., (a) electron and (b-h) ions density profiles along with standard deviation bars as a function of altitude (150–450 km) for pre-, during and post- DSW event (black, red, and magenta colors, respectively) periods.

| Species | $[e^-]_{max.}$ | | $[NO^+]_{max.}$ | | $[O_2^+]_{max.}$ | | $[CO_2^+]_{max.}$ | | $[C^+]_{max.}$ | | $[N^+]_{max.}$ | | $[O^+]_{max.}$ | | $[OH^+]_{max.}$ | |
|---|---|---|---|---|---|---|---|---|---|---|---|---|---|---|---|---|
| | Alt. (km) | Dens. (cm$^{-3}$) | Alt. (km) | Dens. (cm$^{-3}$) | Alt. (km) | Dens. (cm$^{-3}$) | Alt. (km) | Dens. (cm$^{-3}$) | Alt. (km) | Dens. (cm$^{-3}$) | Alt. (km) | Dens. (cm$^{-3}$) | Alt. (km) | Dens. (cm$^{-3}$) | Alt. (km) | Dens. (cm$^{-3}$) |
| Pre-Event | 200 | 210 | 200 | 19.7 | 200 | 125 | 215 | 1.18 | 195 | 1.21 | 195 | 0.67 | 195 | 13.5 | 195 | 6.5 |
| DSW-Event | 250 | 461 | 235 | 26.5 | 235 | 260 | 235 | 1.7 | 235 | 3.5 | 245 | 2.2 | 245 | 78.2 | 235 | 20.5 |
| | | | 280 | 37.4 | 280 | 306 | | | 280 | 3.2 | 280 | 2.0 | 280 | 91.3 | 280 | 27 |
| Post-Event | 200 | 222 | 195 | 21 | 185 | 199 | 210 | 1.04 | 205 | 1.0 | 195 | 0.23 | 195 | 9.2 | 195 | 5.0 |

**Table 1.** The maximum (max.) average magnitude of electron and ions (NO$^+$, O$_2^+$, CO$_2^+$, C$^+$, N$^+$, O$^+$, and OH$^+$) densities in the Martian nightside ionosphere during the pre-, DSW, and post-event periods. The abbreviations (Dens., and Alt.) used here, represent the maximum density and altitude, respectively.

## 4. Discussion

The present investigation shows the impact of solar wind disappearance event on the Martian nightside ionosphere and explores an unusual case of the Mars-solar wind interaction. During the DSW event, we found an enhancement in the electron and ion densities in the nightside ionosphere. The maximum enhancement in electron density is ~2.5 times, whereas, for ions, it is >10 times, respectively, compared to their typical average quiet-time periods (Fig. 2) at a given altitude.

The Martian ionosphere vertical structure primarily depends on the pressure balance between the solar wind and ionosphere (Nagy et al., 2004; Sánchez-Cano et al., 2021). Typically, the solar wind dynamic pressure is ~0.4 × $10^{-9}$ Pa at Mars (Liu et al., 2021), comparable or higher than ionospheric thermal pressure (Nagy et al., 2004; Ma et al., 2008). The higher dynamic pressure, compresses the ionosphere, allowing the penetration of the draped IMF around the planet (Dubinin et al., 2023; Girazian et al., 2019; Tanaka, 1993; Xu et al., 2019). This can affect the formation of both dayside and nightside ionosphere (Shinagawa, 1996; Tanaka et al., 1998) and the pressure balance between the ionosphere and solar wind. However, the remarkable contrast between the extremely low solar wind dynamic pressure (~1.4 × $10^{-11}$ Pa) during the DSW event and the relatively higher nightside ionospheric thermal pressure (~$10^{-10}$ Pa; Fig. 7 of Halekas et al., 2023), nearly by an order of magnitude, triggers an unstable ionospheric state. This condition results in an upward expansion from the lower to the topside ionosphere, likely in the direct response to an unbalanced pressure state or serving as a mechanism for the ionosphere to reach a new state of equilibrium (Flynn et al., 2017; Ma et al., 2014; Qin et al., 2022). Subsequently, it leads to the enhancement in the electron and ions densities at the top nightside ionosphere. These observations are supported by previous studies (Diéval et al., 2014; Girazian et al., 2019; Lillis and Brain, 2013), which have demonstrated the direct influence of solar wind dynamic pressure on the nightside ionosphere. Moreover, the nightside ionosphere is found to be depleted during the high solar wind dynamic pressure periods (ICMEs and CIRs) (Girazian et al., 2019; Krishnaprasad et al., 2019; Ram et al., 2023, 2024; Thampi et al., 2021). In addition, the study by Girazian et al. (2019) showed that the higher dynamic pressure leads to an ionospheric compression and inhibit the day-to-night plasma transport, causes depleted nightside ionosphere. However, in very recent work of Halekas et al. (2023) and Fowler et al. (2024), they demonstrated that during an extremely low solar wind dynamic pressure period, the magnetospheric-ionospheric boundaries expanded by thousands of kilometers on the dayside. The expansion in the dayside ionosphere, therefore, could enhance the day-to-night plasma transport due to pressure gradient (Adams et al., 2018; Cao et al., 2019; Girazian et al., 2017) and may be a contributing factor to enhance the nightside ionospheric densities (Cui et al., 2015; Girazian et al., 2017; González-Galindo et al., 2013; Qin et al., 2022).

In order to depict the overall scenario of the Martian ionosphere for pre-, during, and post-DSW event periods, a schematic diagram is provided in Fig. 3. The top panel (Fig. 3a) shows the location of bow shock (thick dotted black curve line), the size of the magnetosphere (dotted green curve line) and ionosphere (brown shades) of Mars during normal solar wind conditions. Fig. 3b illustrates the impact of the DSW event on the location of the bow shock and the size of the magnetosphere-ionosphere system. This schematic diagram vividly portrays the expansion of bow shock, magnetosphere, and particularly the Martian ionosphere, which is evident from Fig. 2. The extent and shape of the terrestrial magnetosphere mainly depend on the solar wind conditions particularly the dynamic pressure (Ingale et al., 2019, Halekas et al., 2023; Sánchez-Cano et al., 2021). The location of these plasma boundaries is found to be significantly affected during the DSW event due to extremely low solar wind dynamic pressure.

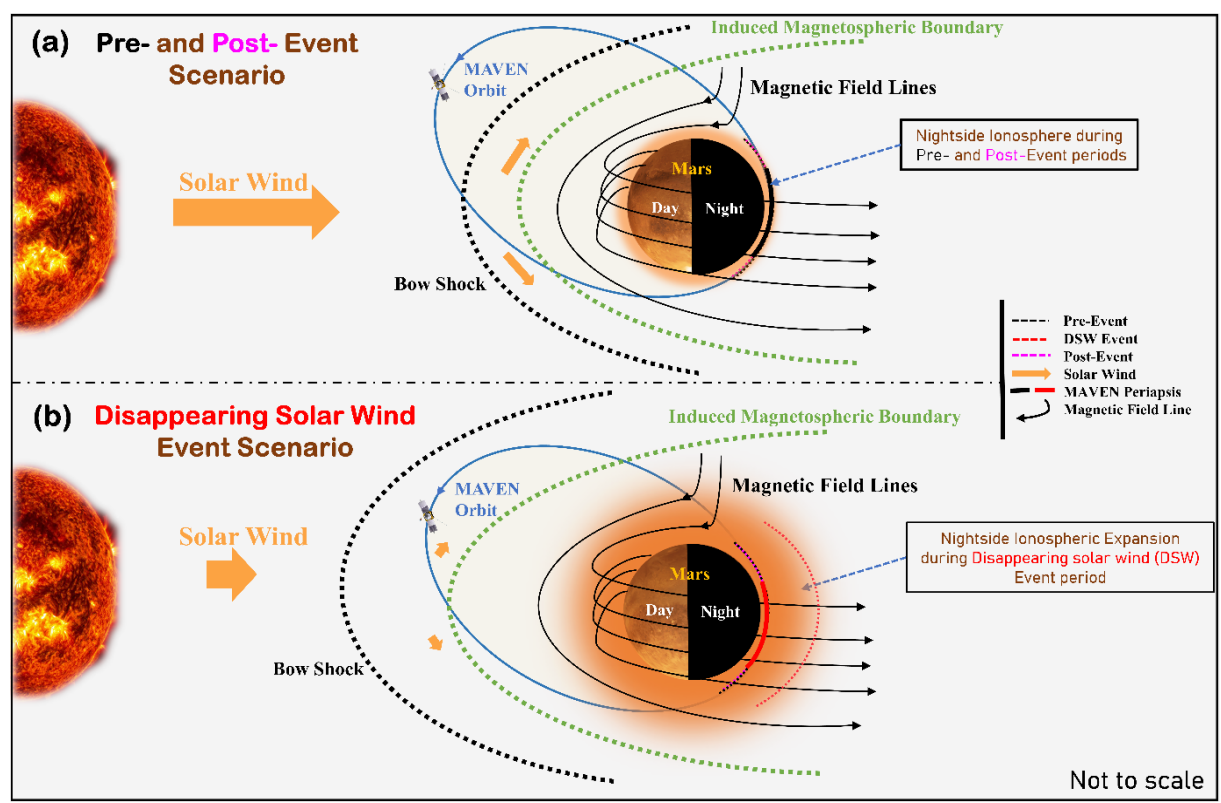

**Fig. 3.** A schematic that portrays the Mars-solar wind interaction. The top panel depicts the Martian ionosphere during pre- and post-event periods, whereas, the bottom panel represents the solar wind disappearance period (26 December 2022). The schematic is not drawn to scale.

Lastly, we envisage some other potential factors to influence the Martian nightside ionosphere such as varying magnetic field topologies (Xu et al., 2019) and the role of strong/weak crustal magnetic field during the DSW event, which remain unexplored so far. The previous results of Li et al. (2023) using a 3D multispecies magnetohydrodynamic (MHD) model, suggested an inhibitory effect of the crustal fields on day-to-night ion transport, leading to reduced ions density in the southern nightside ionosphere. However, during the DSW event, we observed enhanced ions density, suggesting a direct influence of contrasting solar wind dynamic and ionospheric thermal pressure along with a potential contribution from day-to-night transport (Cao et al., 2019). Additionally, the DSW event could trigger an expansion of the cross-terminator crustal field loops, significantly altering the nightside ionosphere. Therefore, this aspect poses an open question and a new avenue for further investigation by the scientific community, to probe the Martian nightside ionosphere, with/without crustal magnetic field (strong/weak) region during the extremely low solar wind density event.

## 5. Summary and Conclusions

We have investigated the variation of the Martian nightside ionospheric species i.e., electron and ions (heavier and lighter) during the periods of a rarest solar wind phenomenon, known as disappearing solar wind (DSW) event. We observed a denser nightside ionosphere during the DSW event, where plasma density increased by a factor of ~2.5 (for electrons) and >10 (for ions) compared to their typical average quiet-time periods. The increased plasma densities at higher altitudes are the result of extremely low solar wind dynamic pressure and day-to-night plasma transport. Thus, this study deepens our understanding of an unusual Mars-solar wind interaction phenomenon and advances us to further study the Mars-like planetary/exoplanetary bodies having magnetized/non-magnetized or hybrid-magnetosphere. In addition, this study provides significant implications for future spacecraft orbit adjustments in the ionospheric region, especially during the periods when the solar wind nearly disappears around the planet. Also, it offered a unique opportunity to understand the Martian ionospheric plasma variation and outflow/escape processes over the dormant phases of the solar wind. Moreover, this study can help us to comprehend the impact of this event on the planetary atmosphere/ionosphere while quantifying the baseline conditions of the magnetosphere and ionosphere system.

## 5. Data Availability

The MAVEN dataset utilized during this work are accessible through the NASA Planetary Data System at https://pds-ppi.igpp.ucla.edu/search/?t=Mars&sc=MAVEN&facet=SPACECRACT _ NAME&depth=1 and MAVEN SDC (https://lasp.colorado.edu/maven/sdc/public/). The MAVEN in-situ datasets using LPW Level 2, version_19, revision_01 (Andersson et al., 2017; Dunn et al., 2023), NGIMS (ions density) Level 2, version_08, revision_01 (Benna & Lyness, 2014), SWIA (solar wind ion) Level 2, version_19, revision_01 (Dunn et al., 2023; Halekas, 2017), and MAG (IMF) Level 2, version_19, revision_01 (Connerney, 2017; Dunn et al., 2023) datasets utilized during this work are accessed through NASA Planetary Data System (PDS) and downloading using the Python Data Analysis and Visualization tool (MAVEN SDC et al., 2020).

## 6. Acknowledgement

We sincerely acknowledge the NASA PDS and MAVEN team especially LPW, MAG, NGIMS, and SWIA instruments team for the datasets. L. Ram acknowledges the fellowship from the Ministry of Education, Government of India for carrying out this research work. This work is also supported by the Department of Space and the Ministry of Education, Government of India.